\documentstyle[preprint,aps]{revtex}   
\begin{document}
\draft
\title{Transition density of diffusion on Sierpinski gasket
and extension of Flory's formula}
\author{Tetsuya Hattori\cite{TH} and Hideo Nakajima\cite{HN}}
\address{Department of Information Science, Utsunomiya University,
Ishii-cho, Utsunomiya 321, Japan}
\date{\today}
\maketitle
\begin{abstract}
Some problems related to the transition density $u(t,x)$
of the diffusion on the Sierpinski gasket are considerd,
based on recent rigorous results and detailed numerical calculations.
The main contents are an extension of Flory's formula for the end-to-end
distance exponent of self-avoiding walks on the fractal spaces, and an
evidence of the oscillatory behavior of $u(t,x)$ on the Sierpinski gasket.
\end{abstract}
In this letter we report our study on the transition density $u(t,x)$
of the diffusion and the random walk on the Sierpinski gasket, based on
recent rigorous results and detailed numerical calculations.
The main contents are an extension of Flory's formula for the end-to-end
distance exponent of self-avoiding walks on the fractal spaces, and an
evidence of the oscillatory behavior of $u(t,x)$ on the Sierpinski gasket.
\par
Recently, rigorous justification of the (symmetric and isotropic) diffusion
on the Sierpinski gasket and analysis of its behavior appeared in
mathematics literatures \cite{SGc,BP}.
Among other results in these studies, we focus on the transition density
$u(t,x)$, the density at point $x$ at time $t>0$, for the diffusion
starting at $t=0$ from the origin of the Sierpinski gasket.
In \cite{BP} $u(t,x)$ is rigorously shown to exist,
and the following form of bound is proved to hold for all $t>0$ and
at any point $x$ on the Sierpinski gasket:
\begin{equation} \label{diff}
f(t,x;C_1,C_2) \le u(t,x) \le f(t,x;C_3,C_4) \,,
\end{equation}
where $C_i$s are some positive constants independent of $t$ and $x$,
and the function $f$ is given by
\begin{equation} \label{BP}
f(t,x;C_1,C_2) = C_1 t^{-d_s/2}
\exp\{ -C_2 (|x| t^{-1/d_w})^{2\eta} \} \,,\ \ \eta=d_w/(2d_w-2)\,.
\end{equation}
The exponents $d_w$ and $d_s$ in (\ref{BP}) are the walk dimension and the
spectral dimension, respectively, whose values for the Sierpinski gasket are
$d_w=\log 5/ \log 2$, and $d_s=2 d_f/d_w= 2\log 3/\log 5$, where
$d_f=\log 3/\log 2$ is the fractal dimension \cite{Mandeletal,HBAHHW}.
The specific form $|x| t^{-1/d_w}$ in (\ref{BP}) implies anomalous diffusion
$<|x(t)|^2> \sim t^{2/d_w}$ (or rather, this relation defines $d_w$).
Note that the value of $\eta$ in (\ref{BP}) cannot be determined
from this relation alone.
Bounds of the form (\ref{diff}) with $\eta$ as in (\ref{BP}) are
mathematically proved to hold also for a wide class of finitely ramified
fractals \cite{nf} (with some generalizations which we will not deal with
here), and even on some infinitely ramified fractals such as
the Sierpinski carpet \cite{SCc}.
The wide applicability of (\ref{BP}) suggests us to take this
formula as one of the basis in the studies of $u(t,x)$.
We consider two problems related to $u(t,x)$.
One is the extension of Flory's formula for the self-avoiding walks (SAW)
to the fractal spaces, and
the other is the oscillatory behavior of $u(t,x)$ on the Sierpinski gasket.
\par
Consider a SAW on a fractal with the fractal dimension $d_f$ and
the spectral dimension $d_s$.
The end-to-end distance exponet $\nu$ is defined by $R(N) \sim N^{\nu}$
($N\gg 1$), where $N$ is the number of steps of a SAW and $R(N)=<|x(N)|>$ is
the average end-to-end distance of $N$-step SAW.
According to the mean-field type arguments for SAW,
Flory's value $\nu_F$ \cite{FG,RTV} for the exponet $\nu$ is obtained by
finding the solution $R=R_F(N)$ which attains the minimum of the
^^ free energy' $-\log u(N,R) + V(N,R)$ for each $N$, where we wrote $u(N,R)$
for an average of the transition density $u(N,x)$ of the simple random walk
over $x$ with $|x|\approx R$, and $V(N,R)= N^2/R^{d_f}$ represents
the volume exclusion effects.
$\nu_F$ is then determined by $R_F(N) \sim N^{\nu_F}$.
The studies that derived (\ref{BP}) for finitely ramified fractals
start with analysis of simple random walks and then reach
the diffusions by taking continuum limits.
Therefore the long time behavior ($N\gg 1$) of the transition density
$u(N,x)$ for a random walk also satisfies (\ref{diff}) with (\ref{BP}).
We use the form (\ref{BP}) for $u(N,R)$ to obtain,
\begin{equation} \label{F}
\nu_F=\nu_F(\eta)=2 \frac{1+\eta/d_w}{d_f+2 \eta}\,,\ \ \eta=d_w/(2d_w-2)\,.
\end{equation}
The argument holds for any network with definite fractal dimension $d_f$
and walk dimension $d_w$.
The value $\eta=d_f/d_s=d_w/2$ was proposed at times when the form of
(\ref{BP}) was not settled, resulting in a
simpler formula $\nu_F(d_w/2)= 3/(d_f+d_w)$ \cite{RTV,fF}.
In \cite{RTV} it was pointed out that there was no justification in this
choice other than simplicity,
and that the problem of the choice of $\eta$ remained open.
A heuristic explanation of the rigorous proof \cite{BP} for the value
$\eta=d_w/(2d_w-2)$ is as follows.
\par
Fix the step $N$ and the distance $R$ such that $R=|x|\gg N^{1/d_w}$
and consider walks of $N$ steps that reach a point at distance $R$;
i.e.\ look at walks going outwards quickly.
Classify the random walk sample paths by the scale $r_0$, such that
for scale $r$ larger (resp., smaller) than $r_0$ the random walker walks
straight (resp., walks randomly; $\Delta N \sim \Delta r^{d_w}$).
A walk specified by the scale $r_0$ passes straight through $R/r_0$ blocks of
scale $r_0$, by definition of $r_0$.
Since it takes steps of order $r_0^{d_w}$ to pass through each block,
we have $N \sim R/r_0 \cdot r_0^{d_w}$, which imply that
the dominant contribution to the quick diffusion specified by $(N,R)$ comes
from the walks with $r_0 \sim (N/R)^{1/(d_w-1)}$.
Each time the walker passes the block straight through he loses probability
by $1/4$, because, at each node, there are $4$ possible directions (i.e.\
the $4$ outmost vertices of the two blocks connected to the node) to go.
The total decay of probability, which gives an estimate of the transition
density is $u(N,R) \sim 4^{-(R/r_0)}$, because the walker passes $R/r_0$
blocks straight through.
Using the estimate for $r_0$ given above, we have
$ -\log p_N(R) \sim (R\, N^{-1/d_w})^{d_w/(d_w-1)} \log 4$,
which implies $\eta=d_w/(2d_w-2)$.
\par
The reason that (\ref{BP}) is to be used for Flory's formula can be seen
from the above argument:
SAW is ^^ pushed outwards' compared to random walks, owing to
self-repulsion or volume exclusion effects.
Therefore, the dominant contribution to SAW comes from those walks
that move quickly away.
The argument given above explains that the value $\eta=d_w/(2d_w-2)$ is the
consequence of the contribution from walks which quickly move away,
hence it is reasonable to use this form in deriving the Flory's formula.
\par
The explicit values of $d_f$ and $d_w$ are known for the Sierpinski gasket
and its natural $d$-dimensional generalizations ($d$SG), constructed by
$(d+1)$-simplex instead of triangle to construct the Sierpinski gasket.
The values are $d_f=\log(d+1)/\log2$ and $d_w=\log(d+3)/\log2$ for $d$SG
\cite{Mandeletal,HBAHHW}, with which $\nu_F=\nu_F(d_w/(2d_w-2))$ can be
calculated from (\ref{F}).
For $2$SG ($=$ Sierpinski gasket) and $3$SG,
the values of $\nu_F$ ($0.8249 \cdots$ and $0.724588 \cdots$, respectively)
are to be compared with the exact values of $\nu$, which are
\begin{equation} \label{nu} \begin{array}{l}
\nu(2SG)=\log 2/ \log \, (7-\sqrt{5})/2 = 0.79862 \cdots \,,
\\ \nu(3SG)=0.67402 \cdots \,.
\end{array} \end{equation}
The value of $\nu(3SG)$ has similar exact expression as that for
$\nu(2SG)$, but in place of integers $7$ and $5$ appear roots of
a $14$-th order algebraic equation \cite{HHK2}.
The values in (\ref{nu}) have been known for some time \cite{Dhar} (see also
\cite{RTV,BAHKS}), and are recently proved rigorously in \cite{HHK,HHK2}.
Flory's formula (\ref{F}) is within 3\% and 8\% precision from
the exact values for $2$SG and $3$SG, respectively.
Flory's formula is known to be numerically very good for SAW on
Euclidean lattices (for a recent nice review of SAW on Euclidean lattices,
see \cite{MS}).
The extended Flory's formula (\ref{F}) which we have is not very bad,
but not very close to the exact values compared to Euclidean cases.
If we put $\eta=1$, the values become closer (in fact, it is close to the
best choice) to the exact results (0.06\% and 3\% deviation for
$2$SG and $3$SG, resp.).
The choice $\eta=d_w/(2d_w-2)$ has the most sound basis (\ref{diff}) and
(\ref{BP}), but the value is better for $2$SG and $3$SG with $\eta=1$.
Deviation of (\ref{F}) from the rigorous and exact results (\ref{nu})
leads our interest to detailed numerical studies of $u(t,x)$.
\par
Two open problems are found in the literatures concerning the detailed
structure of $u(t,x)$.
One problem is the value of $\eta$; the results given in \cite{Guyer} agree
with (\ref{BP}), while those in \cite{BWSP} claim the value $\eta=d_w/2$.
The other problem is the observation in \cite{BP,FS}
that there are ^^ oscillations' in $u(t,x)$.
\par
To perform numerical calculations, we regard $u(t,x)$ as the electric charge
density of a point $x$ at time $t$, and reformulate the problem in terms of
the impedance circuits.
We consider a $d$-dimensional Sierpinski gakset $d$SG.
The corresponding electrical circuit has impedance distribution on the
gasket, and also impedance between the gasket and the ground.
By symmetry and star--triangle ($Y-\Delta$) type relations, a unit block of
the gasket (a sub-circuit of $d$-dimensional simplex with side length $1$)
can effectively be represented by a device with $d+1$ terminals, each
connected by an impedance $a(s)$ to the center point of the simplex,
to which the ground is connected by an impedance $b(s)$.
$s$ is the dual variable to $t$ in the Laplace transform.
The self-similarity of the diffusion implies the scaling behavior
$B\,u(L t,2 x)=u(t,x)$ \cite{SGc,BP,Mandeletal,HBAHHW},
where we put $L=d+3=2^{d_w}$ and $B=d+1=2^{d_f}$.
Using the scaling behavior, the self-similarity of the gasket,
and the similarity among $d+1$ terminals of a block simplex,
together with star--triangle type relations, we find
\begin{equation} \label{recdif}
(a(L s)\, L/B ,\,g(L s))=W(a(s),\,g(s)) \,,
\end{equation}
where $g(s)=2a(s)/b(s)$, and $W(x,y)=(x (y+L)/(y+B), y (y+L))$.
We also find $g(0)=0$.
$g'(0)$ and $a(0)$ determine the normalization of $u$ and $t$.
We focus on the normalization independent quantities such as exponents
and oscillations.  By fixing $g'(0)$ and $a(0)$,
the solution of the functional equation (\ref{recdif}) is determined uniquely.
The equation for $g(s)$ is known as Schr\"{o}der's functional equation.
The existence of the solution has been studied \cite{SM},
but its detailed behavior seems to be unknown.
\par
We define two asymptotic functions $\displaystyle
C(s)=s^{1-d_s/2} \lim_{n\to\infty} (L/B)^n a(L^n s) $ and $\displaystyle
k(s)=s^{-1/d_w} \lim_{n\to\infty} 2^{-n} \log g(L^n s) $, for $s>0$.
These functions are periodic in $\log s$ with period $\log L$,
hence can be expanded in Fourier series:
\begin{equation} \label{Fourier} \begin{array}{l} \displaystyle
C(s)=c_0 + \sum_{n=1}^{\infty} c_n \sin(2\pi n \log_L s + \phi_n) \,,
\\ \displaystyle
k(s)=k_0 + \sum_{n=1}^{\infty} k_n \sin(2\pi n \log_L s + \phi'_n) \,.
\end{array} \end{equation}
We numerically obtained by double precision FORTRAN calculations for $d=2$,
the Sierpisnki gasket; $c_1/c_0=1.21929438 \times 10^{-5}$, and $c_2/c_1=3.68
\times 10^{-6}$, for $C(s)$, and $k_1/k_0=1.5264191 \times 10^{-6}$,
and $k_2/k_1=5.6 \times 10^{-7}$, for $k(s)$.
The results show a strong hierarchy of coefficients, such that the higher
frequency components have exponentially small
values ($c_n =O(10^{-5n})$, $k_n =O(10^{-6n})$).
Note that we have small but non-zero numbers of $O(10^{-6})$
out of equation (\ref{recdif}) with $O(1)$ coefficients,
which is potentially an interesting phenomena.
We performed the numerical calculations up to $d=10$,
and obtained qualitatively similar behavior,
with somewhat larger amplitudes of oscillations for larger $d$.
For precision check, we performed the calculations for $d=1$, corresponding
to the diffusion on a line, and obtained the correct constant values
$C(s)=2^{1/2}$ and $k(s)=2^{1/2}$, within error $10^{-14}$.
The peak values of $k(s)$ are consistent with the numerically
obtained values in \cite{BP}.
\par
We can calculate the Laplace transform $\tilde{u}(s,x)$ of the density
using the impedances.
For the transition density at the origin, we have
$\tilde{u}(s,0)=a_{\infty}(s)/2$, where $a_{\infty}(s)=C(s) \, s^{d_s/2-1}$
is the impedance corresponding to a simplex of ^^ infinite' size.
Using (\ref{Fourier}), we can evaluate the inverse Laplace transformation
of $\tilde{u}(s,0)$, term by term in the Fourier series, using change of
contours.  We have
\[ u(t,0)= N\, t^{-d_s/2}
[c_0 p_0 + \sum_{n=1}^{\infty} c_n p_n \cos(2\pi n \log_L t-\psi_n)] \,, \]
with $c_n$ as in (\ref{Fourier}) and
$p_n = |\Gamma(2^{-1} d_s+\pi^{-1} \Omega n i)| 2^{-1/2}
(\cosh(2 \Omega n)-\cos(\pi d_s))^{1/2}$ for $n\ge 0$,
where $\Omega=2\pi^2/\log L$.
$N$ is a normalization constant.
Thus the oscillation in $C(s)$ explains that in $u(t,0)$.
\par
We parametrize $u(t,x)=f(t,x;C_1(t),C_2(t,x))$ with $f$ as in (\ref{BP}),
and consider the oscillations in $C_i$s.
Using the values given below (\ref{Fourier}) for $c_n$s, we have, for $d=2$,
$C_1(t) =t^{d_s/2} u(t,0)
= C_{10}+ C_{11} \cos(2\pi \log_L t - \psi_1) + \cdots$,
with $C_{11}/C_{10}=8.0964779 \times 10^{-3}$.
We performed numerical Laplace inverse transformation of $\tilde{u}(s,0)$
and obtained a consistent value.
The Laplace transform of the density $u(t,1)$ at a vertex of a unit simplex
is given by $\tilde{u}(s,1)/\tilde{u}(s,0)
=(1+g(s)/(2d)) (1-a(s)/a_{\infty}(s)) - a(s)/(d a_{\infty}(s))$.
$C_2(t)=C_2(t,1)$ is then given by
$C_2(t)=t^{2\eta/d_w} \log (u(t,0)/u(t,1))$.
To calculate the inverse Laplace transforms for small $t$ ($t \le 10^{-2}$),
we find $s=s_0>0$ which gives minimum of $\tilde{u}(s,1) \exp(st)$ (as we do
in the steepest descent method of complex contour integration), and
numerically evaluate the contour integration with the contour $\Re(s)=s_0$.
For larger $t$ we use the contour $\Re(s)=1/t$.
(The details of the numerical calculations will be reported elsewhere.)
Figs.~\ref{f1} and \ref{f2} show $C_2(t)$ for $d=2$ as a function of $\ln t$.
Fig.~\ref{f2} shows the small oscillation ($O(10^{-6})$ in amplitude)
in $C_2(t)$ for very small $t$.
We performed similar calculations for the exactly solvable $d=1$ case, and
checked that the error in the range of Fig.~\ref{f2} is $O(10^{-11})$.
Small $t$ corresponds to large $s$ in the Laplace transform, where we have an
asymptotic formula $\tilde{u}(s,1)/\tilde{u}(s,0)\sim \exp(-s^{1/d_w}k(s))$.
Thus the oscillation in $k(s)$ explains that in $C_2(t)$ for very small $t$.
\par
{} From Fig.~\ref{f1} we see that (besides the tiny oscillation)
$C_2(t)$ is flat for small $t$.
This is consistent with the fact \cite{BP} that the value $\eta=d_w/(2d_w-2)$
explains the asymptotic behavior of $u(t,x)$ as $t\to 0$.
For larger $t$, $C_2(t)$ is decreasing with a significant size of oscillation.
If we try to explain this decrease in terms of ^^ effective (dynamical)'
changes in the value of $\eta$,
i.e. keep $C_2$ constant and let $\eta$ change as $\eta=\eta_{eff}(t)$,
we see that the effective value $\eta_{eff}(t)$ increases as $t$ is increased.
Our data are in favor of the argument in \cite{Guyer} that
$\eta_{eff}(t)=d_w/2$ \cite{BWSP} may be effectively good for $t=O(1)$, while
$\eta=d_w/(2d_w-2)$ is good for $t \ll 1$.
Note also that $d_w/(2d_w-2)<1<d_w/2$,
where $\eta=1$ gives good $\nu_F(\eta)$.
The oscillatory behavior of the data prevents us from obtaining precise
results on the value of $\eta_{eff}$.
In contrast to the clarity in the meaning of the value $\eta=d_w/(2d_w-2)$,
theoretical basis for $\eta_{eff}$ is still unclear, which has to be settled
before we can be conclusive about its value and implications.
\par
The authors would like to thank Prof.\ M.\ T.\ Barlow and Prof.\ S.\ Kusuoka
for discussions, and Prof.\ N.\ Yanagihara for bringing references on
Schr\"{o}der's functional equations.
The research of T.\ Hattori is supported in part by a Grant-in-Aid for General
Scienctific Research from the Ministry of Education, Science and Culture.

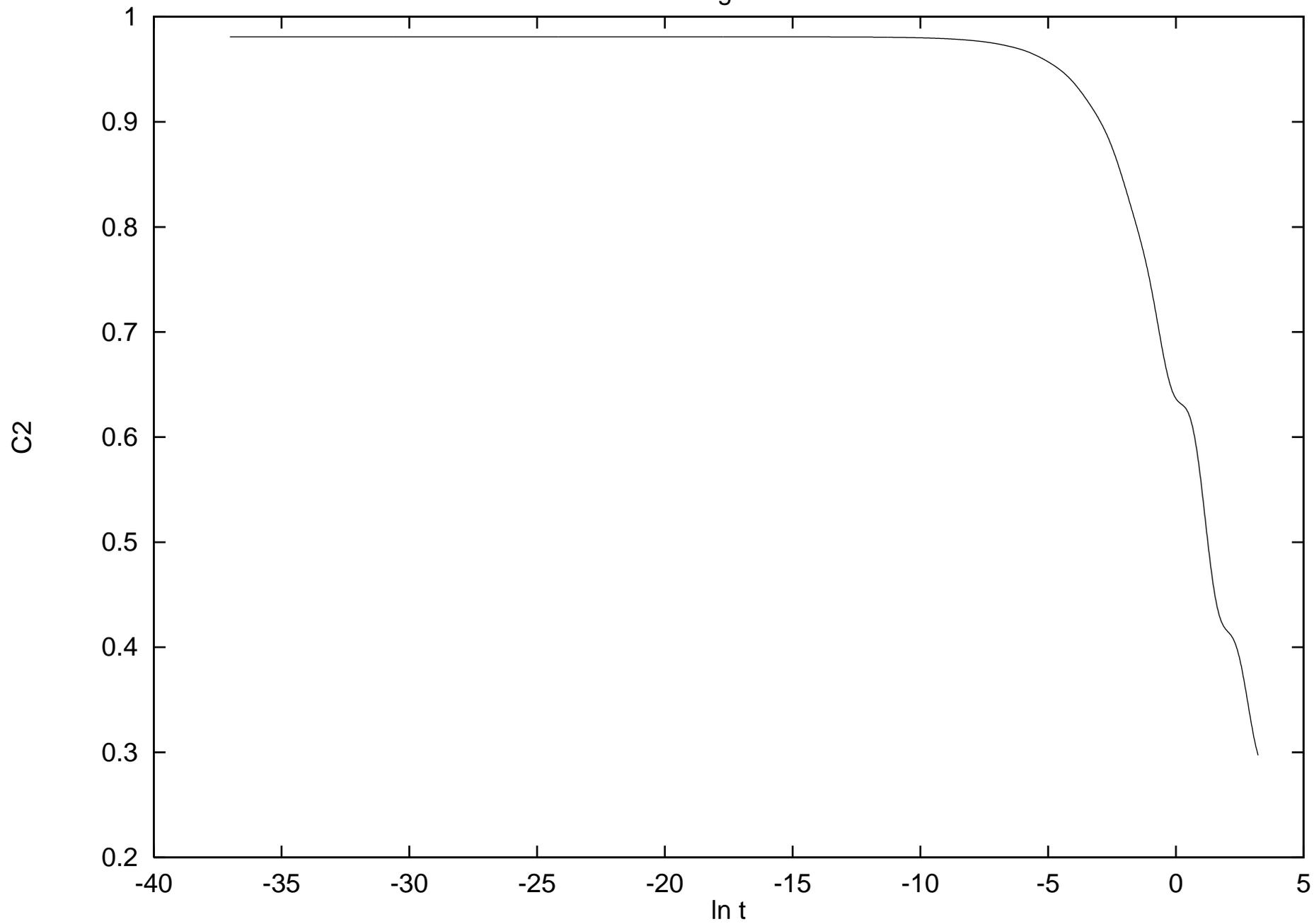
\begin{figure}
\caption{$C_2(t)=t^{2\eta/d_w} \log (u(t,0)/u(t,1))$.} \label{f1}
\end{figure}
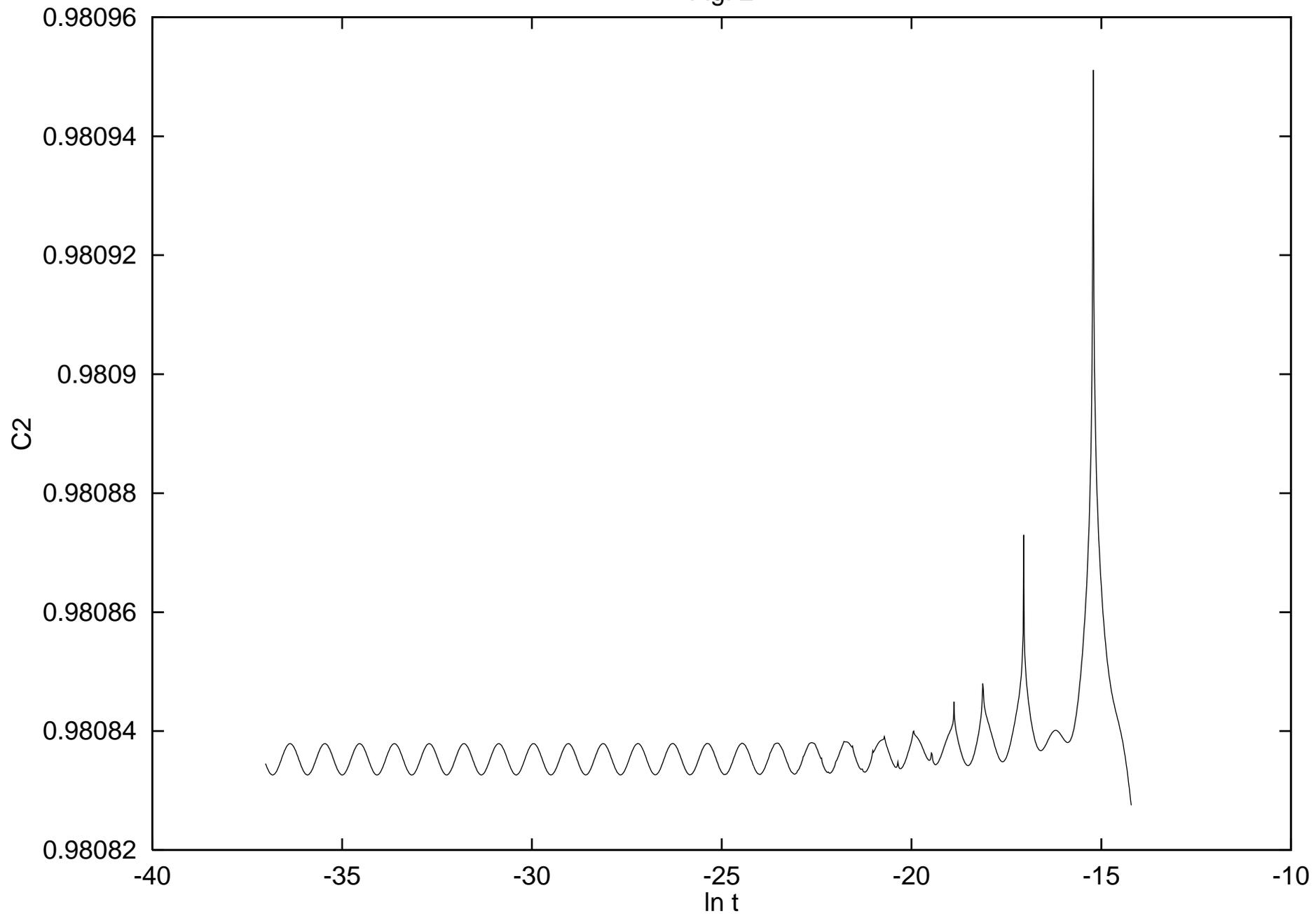
\begin{figure}
\caption{Oscillation of $C_2(t)$ for small $t$.} \label{f2}
\end{figure}
\end{document}